\documentclass[aps,prl,twocolumn,superscriptaddress,floatfix,longbibliography,nofootinbib]{revtex4-2}

\usepackage{amsmath,amssymb}
\usepackage{graphicx,xcolor}% Include figure files
\usepackage{dcolumn}% Align table columns on decimal point
\usepackage{bm}% bold math
\usepackage{hyperref}% add hypertext capabilities
\usepackage{graphicx}
\usepackage{subcaption}
\usepackage{caption}
\begin{document}

%\title{Opinion Waves in the Voter Model with Latency}
\title{Spontaneous Opinion Swings in the Voter Model with Latency}

\author{Giovanni Palermo}
\email{giovanni.palermo@cref.it}
 \affiliation{Physics Department, Sapienza University of Rome, 00185 Rome, Italy}
 \affiliation{Enrico Fermi Research Center, 00184 Rome, Italy}
\author{Anna Mancini}
 \affiliation{Physics Department and INFN, University of Rome Tor Vergata, 00133 Rome, Italy}
 \affiliation{Enrico Fermi Research Center, 00184 Rome, Italy}
\author{Antonio Desiderio}
 \affiliation{Physics Department and INFN, University of Rome Tor Vergata, 00133 Rome, Italy}
 \affiliation{Enrico Fermi Research Center, 00184 Rome, Italy}
\author{Riccardo Di Clemente}
\affiliation{Network Science Institute, Northeastern University London, London E1W1LP, United Kingdom}
\author{Giulio Cimini}
 \affiliation{Physics Department and INFN, University of Rome Tor Vergata, 00133 Rome, Italy}
 \affiliation{Enrico Fermi Research Center, 00184 Rome, Italy}

%\date{\today}

\begin{abstract}
The cognitive process of opinion formation is often characterized by stubbornness or resistance of agents to changes of opinion. 
To capture such a feature we introduce a constant latency time in the standard voter model of opinion dynamics: after switching opinion, an agent must keep it for a while. 
This seemingly simple modification drastically changes the stochastic diffusive behavior of the original model, leading to deterministic dynamical oscillations in the average opinion of the agents. We explain the origin of the oscillations and develop a mathematical formulation of the dynamics that is confirmed by extensive numerical simulations. 
We further characterize the rich phase space of the model and its asymptotic behavior. 
Our work offers insights into understanding and modeling opinion swings in diverse social contexts.
\end{abstract}

\maketitle

%\section{\label{intro}Introduction}

%general intro on binary opinion models + the voter model
Binary-choice opinion formation models are popular in the statistical physics community to describe the evolution of opinions within a population of interacting agents \cite{castellano2009,galam2012,redner2019,peralta2022}. 
As for spin systems, in these models the opinion of an agent can take one of two possible values and is influenced by the opinions of other agents through simple dynamic rules, which are iterated until some (ordered or disordered) stable state is reached---either consensus or coexistence of different opinions. 
%[in the economics literature, such peer-pressure-influenced choices are called binary choices with externalities \cite{schelling1973}]
The \emph{voter model} \cite{clifford1973,holley1975} is a paradigmatic framework where an agent changes opinion by selecting one of a randomly chosen neighbor, thus mimicking the processes of conformation and peer influence on the individual's mind \cite{schelling1973,asch2003}.
The model has been studied extensively and has found applications in many fields; moreover, it is one of the very few non-equilibrium stochastic processes that can be solved exactly in any dimension \cite{redner2001}. 

%realistic voter model, variants and stubborn agents, non-markovian dyn, latent
Models of opinion dynamics provide useful tools to probe socio-political scenarios, test descriptive theories of collective behavior for consistency, and explore emergent phenomena \cite{holme2015}. 
However, they are based on simplified hypotheses of human interaction that neglect a lot of psychological and social factors influencing the decisions of individuals. 
A lot of effort has thus been put into extending the basic models by incorporating more realistic aspects of opinion-making  \cite{redner2019,peralta2022}, as well as on calibrating model features on empirical data \cite{galesic2019}. 
Some models try to incorporate a sort of reluctance of the agents to change opinion, often observed in empirical studies \cite{moscovici1980,moscovici1985}. 
The extreme example in this direction is the voter model with ``zealots'' \cite{mobilia2007,redner2019}, where the presence of stubborn agents who do not change opinion at all determines the route to the consensus state \cite{galam2007,xie2011}. 
Other approaches consider memory-dependent rules for opinion changes. In the model by Stark \emph{et al.} \cite{stark2008}, changes of opinions are subject to inertia (the longer an agent maintains her opinion, the less likely she will change it), which can speed up or slow down the reaching of consensus. 
A similar behavior is observed in the model by Wang \emph{et al.} \cite{wang2014}, characterized by a freezing period (agents who changed opinions are less likely to change it in the short run). This kind of mechanism, representing a cost or restrictions associated with opinion switching, was first proposed in the ``latent'' voter model by Lambiotte \emph{et al.} \cite{lambiotte2009}: after changing opinion, an agent sticks to it for a stochastic latency period. This additional rule drives the system away from the consensus state, as the two opinions coexist in the system for very long times. 
The latent voter model has been further studied in the limit of small, exponentially distributed latency times \cite{huo2018latent} and, for slightly different dynamical rules, in the context of differential latencies for the two opinions \cite{montesdeoca2011,scheidler2011}.
%questi sono majority-rule

\begin{figure*} % Place the figure* at the top of the page
    \begin{subfigure}[b]{0.49\textwidth}
        \centering
        \includegraphics[width=\linewidth]{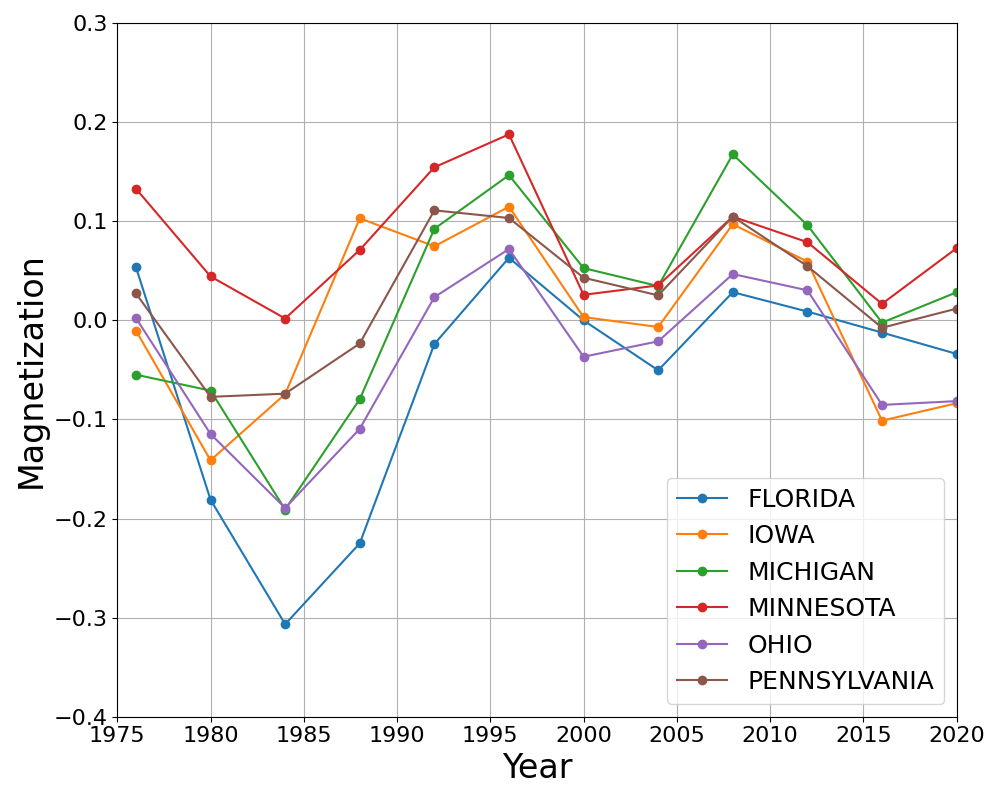}
                \caption{}
                \label{issues}
    \end{subfigure}
    \hfill
    \begin{subfigure}[b]{0.49\textwidth}
        \centering
        \includegraphics[width=\linewidth]{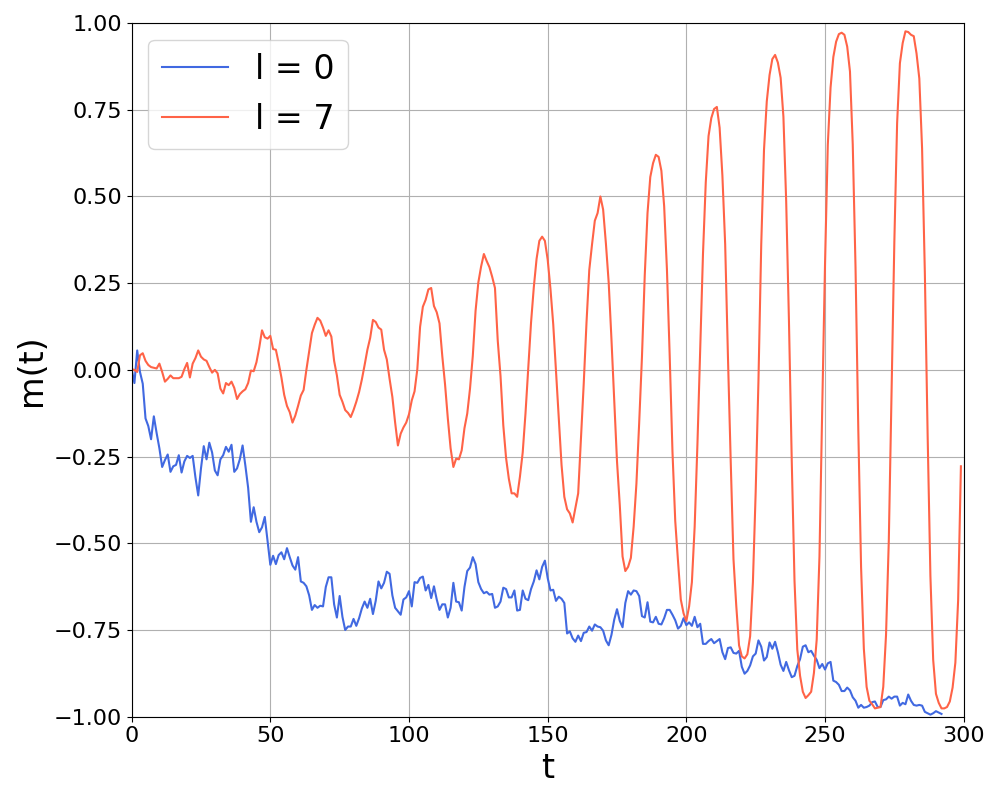}
                \caption{}
                \label{comparison}
    \end{subfigure}
    % \hfill
        % \vspace{20pt} % Adjust vertical space between rows
    
    \begin{subfigure}[b]{0.49\textwidth}
        \centering
        \includegraphics[width=\linewidth]{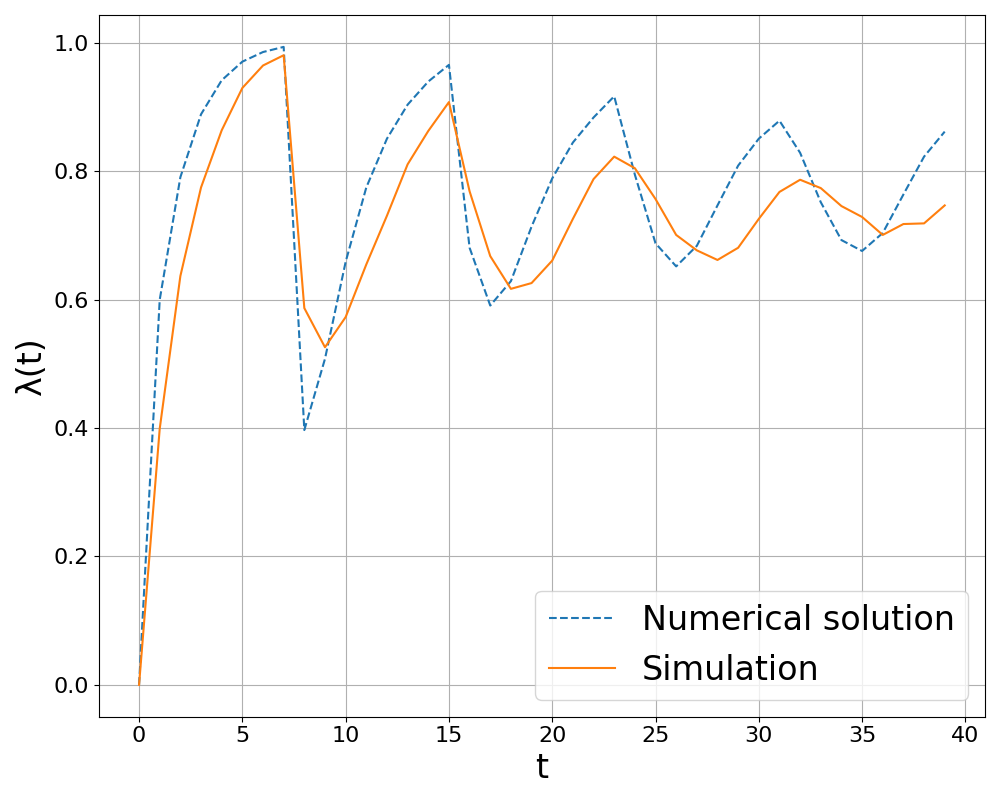}
                \caption{}
                \label{lat30}
    \end{subfigure}
        \hfill
        \begin{subfigure}[b]{0.49\textwidth}
        \centering
        \includegraphics[width=\linewidth]{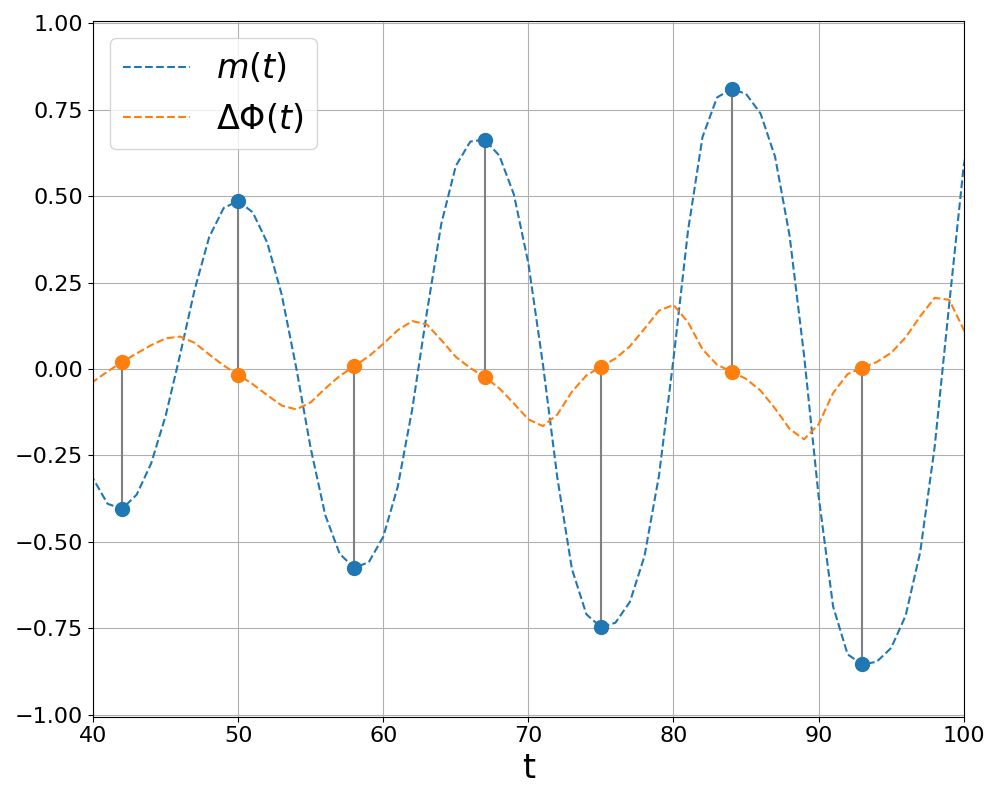}
                \caption{}
                \label{counterphase}
    \end{subfigure}
    \captionsetup{justification=raggedright,singlelinecheck=false}
    \caption{(a) Results of US presidential elections in a sample of \emph{swing states} \cite{DVN/42MVDX_2017}. The dynamic shows oscillating behavior with a period set by the occurrence of elections every 4 years. (b) Evolution of the magnetization $m(t)$ for a single realization of the ordinary voter model and of the voter model with latency, for a population of $N=1000$ agents. Note how the original model can be recovered as a special case of the LVM by setting a latency time $l=0$. 
    (c) Simulation results and numerical solution of eq. \eqref{eq_lam} for the fraction of agents in the latency state $\lambda(t)$, with latency time $l=7$ and $N=1000$ agents (d) The numerical solution with $l=5$ for $m(t)$ and $\Delta \phi(t)$ shows that these quantities oscillate with a quarter-period shift (markers on peaks of $m$ correspond to zeros of $\Delta \phi$).}
\label{fig1}
\end{figure*}

%validation on US data, what about opinion swings?
Concerning model validation, election data represent an ideal test ground, particularly to identify which mechanisms are the most relevant in emulating the social behavior of humans \cite{fernandezgracia2014,braha2017,kononovicius2017}. A handful of studies managed to reproduce some statistical regularities of how votes in US presidential elections (a natural binary opinion setup) are distributed in the population. 
Fernandez-Garcia \emph{et al.} \cite{fernandezgracia2014} were able to capture vote-share fluctuations across counties and long-range spatial correlations, using a noisy voter model with the addition of recurrent mobility of agents. 
Braha \emph{et al.} \cite{braha2017} extended the voter model with opinion leaders and external influence, reproducing geographical patterns of vote-share distribution and social influence. 
A recurring feature observed in electoral data is the presence of opinion waves, especially for the so-called \emph{swing states} \cite{ansolabhere2006}, with temporal patterns determined by the regular occurrence of elections (see Figure \ref{issues} and Supplementary Information S1). 
%\url{https://library.cqpress.com/elections/}
The presence of seemingly regular oscillations of opinions is also widely observed for census data on ``partisan'' issues, such as being pro or against the death penalty or tax increase, or for trend reversals in fashion, and many more \cite{atkinson2021dynamics}. 

%our contribution and motivation
For the first time to the best of our knowledge, in this work we report the emergence of regular opinion swings in binary models of opinion formation.
We do so using the framework of the latent voter model, introducing a homogeneous latency time for each agent after a change of opinion. 
The assumption of the latency time is realistic for partisan issues or in political elections, where a voter switching party will hardly rethink her decision soon. 
Additionally, in these situations a natural time scale exists (the regular occurrence of elections) and sets the same latency time for each agent. 
The introduction of a homogeneous latency time in the voter model leads to the emergence of oscillations in the average opinion, which are not a consequence of stochastic fluctuations, but arise from a deterministic drift due to the non-Markovianity of the update rule. 
%We explain the reason behind the occurrence of such opinion waves and provide a mathematical description of the model in the mean-field regime. 
%Through extensive numerical simulations, we characterize the rich phase space of the model, featuring a phase transition from a constantly swinging opinion to full consensus for finite systems, whereas, in the thermodynamic limit, opinions keep swinging forever.

%\section{\label{model}The Voter Model with Latency}

%We propose a simple modification to the voter model (from now on VM) on a network.
\medskip
\paragraph*{Voter Model with Latency} (LVM). 
We consider a population of $N$ agents, located on the nodes of a fully connected graph. 
Each agent $i$ is assigned with a spin taking values $s_i\in\{\pm1\}$, representing her binary-state opinion. 
At each time step of the dynamics, each agent changes opinion by adopting that of a randomly chosen neighbor. 
With respect to the standard voter model, here we introduce a constant \emph{latency time} $l$: 
when an agent changes opinion she becomes inactive and cannot change it further for the subsequent $l$ time steps (however, she can still influence others).\footnote{Such a modification was originally introduced by Lambiotte \emph{et al.} \cite{lambiotte2009}, yet in a different fashion as an agent in the latent state is reactivated with a fixed probability at each step. Here instead the agent exits the latent state after a fixed number of time steps.} 
The LVM thus formulated has a completely different behavior than the original voter dynamics (see Figure \ref{comparison}). 
In the latter, the average opinion of the population -- given by the magnetization $m=\frac{1}{N}\sum_{i=1}^{N}s_i$ -- follows a stochastic evolution leading to consensus in a time that scales as $\sqrt{N}$. 
The LVM is instead characterized by a predictable evolution, in which $m(t)$ oscillates with a specific frequency and an amplitude that quickly approaches a value close to $1$.\footnote{This behavior is also totally different from what observed in \cite{lambiotte2009}, where the stochastic latency leads to a diffusive dynamics that remains around $m=0$.} 
As we will see, such oscillations can eventually reach the consensus state $|m|=1$ as a finite-size effect.

%\subsection{Model dynamics}
\medskip
\paragraph*{Model dynamics.}
We denote by $N_+(t)$ and $N_-(t)$ the number of agents with spin up and down respectively, while $L_+(t)$ and $L_-(t)$ are the agents with spin up and down who are in the latency state. To characterize the evolution of the system, besides the magnetization $m(t)=[N_+(t)-N_-(t)]/N$ we use the fraction of agents in the latent state with positive and negative spin, respectively $\lambda_+(t)=L_+(t)/N$ and $\lambda_-(t)=L_-(t)/N$. Thus $\lambda(t)=\lambda_+(t)+\lambda_-(t)$ is the total fraction of agents in the latent state (see Supplementary Information S2 for a plot of all these variables). Unless differently stated, we consider initial conditions of zero magnetization, $m(0)=0$, and no agent in the latency state, $\lambda(0)=0$.

We start by writing down the equations for the time evolution of such quantities in the mean-field regime. We set the time scale so that we have $N$ opinion updates in each time step of the dynamics (one for each agent). Denote by $\phi_{-+}(t)$ the probability of a spin flip (\emph{i.e.}, that an agent changes her opinion) from $-1$ to $+1$ at time $t$, and by $\phi_{+-}(t)$ the probability of a spin flip from $+1$ to $-1$. 
The expected value of $\lambda(t)$ is thus given by the sum of all the flip probabilities of a single agent in the past $l$ steps:
\begin{equation}
E[\lambda(t)]=\sum_{\tau=0}^{l}\left[\phi_{-+}(t-\tau)+\phi_{+-}(t-\tau)\right],
\end{equation}
where the two terms correspond to the expected values of $\lambda_+(t)$ and $\lambda_-(t)$:
\begin{equation}
E[\lambda_\pm(t)]=\sum_{\tau=0}^{l}\phi_{\mp\pm}(t-\tau).
\end{equation}
In these equations, we only sum all flips that occurred in the last $l$ steps, since agents exist from latency afterwards. 
On the contrary, to compute the magnetization we need to account for all the flips of an agent that occurred in the evolution of the system up to time $t$:
\begin{equation}
E[m(t)]=2\sum_{\tau=0}^{t}\left[\phi_{-+}(\tau)-\phi_{+-}(\tau)\right].
\label{eq:m}
\end{equation}

We now drop the notation $E[...]$ and switch from discrete to continuous time. We can connect the equations by showing how the spin-flip probabilities depend on these variables. To compute $\phi_{\mp\pm}$ we have to consider the probabilities of three occurrences: pick an agent who is not in the latency state; the selected agent has spin $\mp1$; pick a neighbor (\emph{i.e.}, a generic agent whatever her latency state) with a spin equal to $\pm1$. Overall we have:
\begin{align}
  \phi_{\mp\pm}(t)    &= [1-\lambda(t)]\left[ \frac{N_\mp(t) -L_\mp(t)}{N-L(t)}\right]\left[\frac{N_\pm(t)}{N}\right] \nonumber \\
   &= \frac{1 \pm m(t)}{2}\left( \frac{1 \mp m(t)}{2} -\lambda_\mp(t) \right).
\label{eq:5}
\end{align}

We can now substitute such expressions into the expected values of $m$ and $\lambda_\pm$, obtaining integral equations whose time derivative leads to the final system of differential equations describing the evolution of the LVM:
\begin{equation}
m'(t)=2\left[\frac{1-m(t)}{2}\lambda_+(t)-\frac{1+m(t)}{2}\lambda_-(t)\right],
\label{eq_m}
\end{equation}
\begin{eqnarray}
\lambda_\pm'(t)&=& \left( \dfrac{1\mp m(t)}{2} \right) \left[\dfrac{1 \pm m(t)}{2} - \lambda_\mp(t)\right] \\ \nonumber
&-& \left( \dfrac{1 \mp m(t-l)}{2} \right) \left[\dfrac{1\pm m(t-l)}{2}-\lambda_\mp(t-l)\right].
\label{eq_lam}
\end{eqnarray}
The system above is a set of nonlinear DDEs (Delay Differential Equations) with constant delay, which cannot be solved analytically due to the nonlinearity of the equations \cite{bellman1963} (see Supplementary Information S3 for an approximate solution for small values of $m$). 

%\subsection{Numerical solution}

As the evolution of $m(t)$ and $\lambda(t)$ ultimately depends only on the flip probabilities $\phi_{\mp\pm}(t)$, we tackle the system of DDEs \eqref{eq_m} and \eqref{eq_lam} using the following iterative method. At each time step $t$: 
$i$) Compute $\phi_{\mp\pm}$ as a function of $m$ and $\lambda_\pm$; 
$ii$) Assess the number of flips as $N \phi_{\mp\pm}$, since $N$ updates occur in a time step; 
$iii$) Update $m$ and $\lambda_\pm$ accordingly, then increase $t$.
These steps are iterated until consensus or manual stop.
The proposed algorithm mimics the simulations of the model, allowing us to obtain a numerical solution for the DDEs that works well in reproducing the peculiar oscillating behavior of $\lambda(t)$ (see Figure \ref{lat30}) and other properties of the system, as we shall see below (see Supplementary Information S4 for more details on the comparison between simulations and numerical solution). 
Additionally, we are able to control for finite-size effects, since the minimum increment is $2/N$ for $m$ and $1/N$ for $\lambda_\pm$, which corresponds to the smallest distance the magnetization can reach from the full consensus state $|m|=1$. 

\begin{figure*} % Place the figure* at the top of the page

    \begin{subfigure}[b]{0.49\textwidth}
        \centering
        \includegraphics[width=\linewidth]{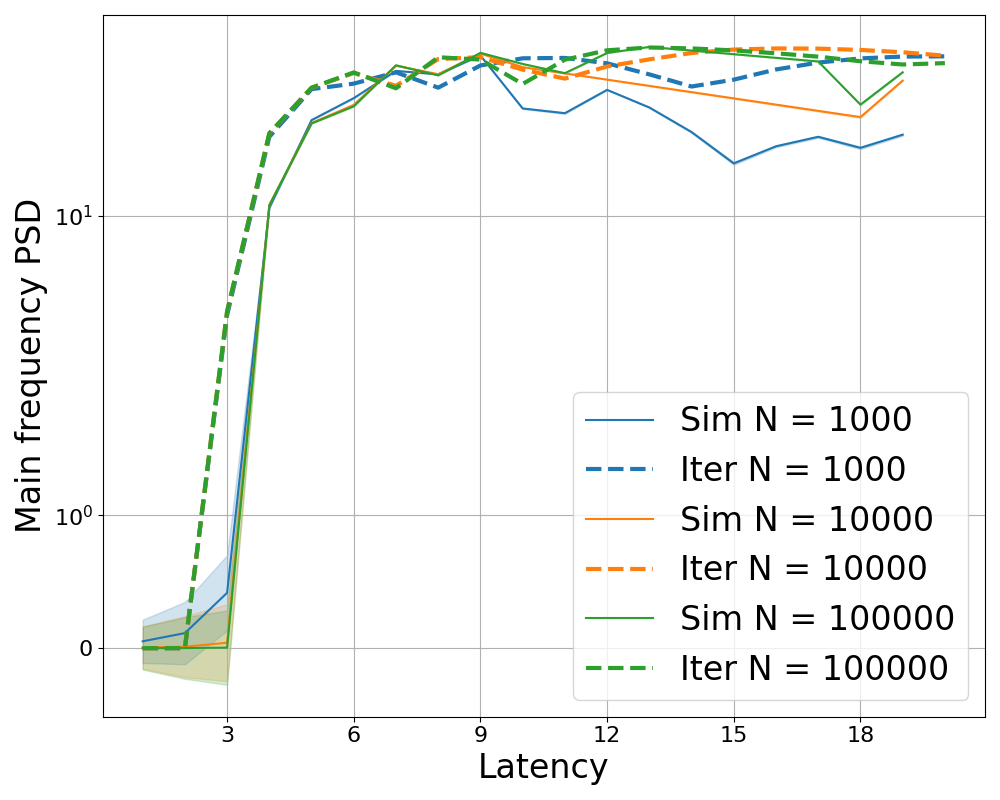}
                \caption{}
                \label{psd_cross}
    \end{subfigure}
    \hfill
    \begin{subfigure}[b]{0.49\textwidth}
        \centering
        \includegraphics[width=\linewidth]{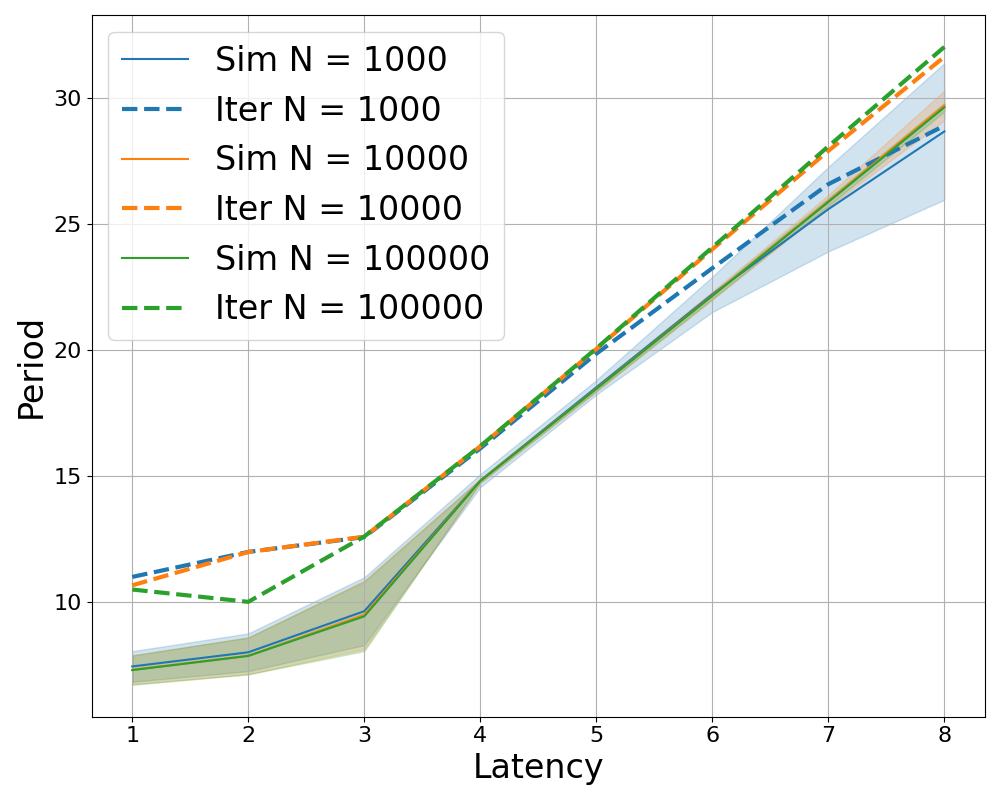}
                \caption{}
                \label{period_cross}
    \end{subfigure}
    % \hfill
        % \vspace{20pt} % Adjust vertical space between rows

    \begin{subfigure}[b]{0.49\textwidth}
        \centering
        \includegraphics[width=\linewidth]{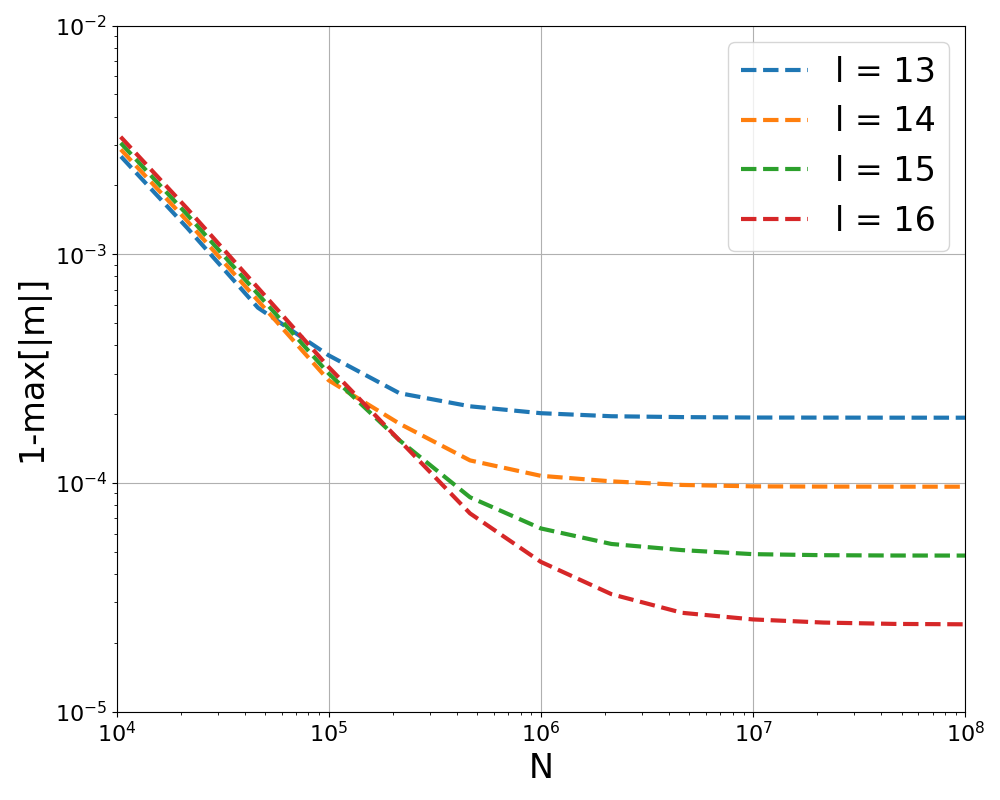}
                \caption{}
                \label{iter_sol}
    \end{subfigure}
         \hfill
        \begin{subfigure}[b]{0.49\textwidth}
        \centering
        \includegraphics[width=\linewidth]{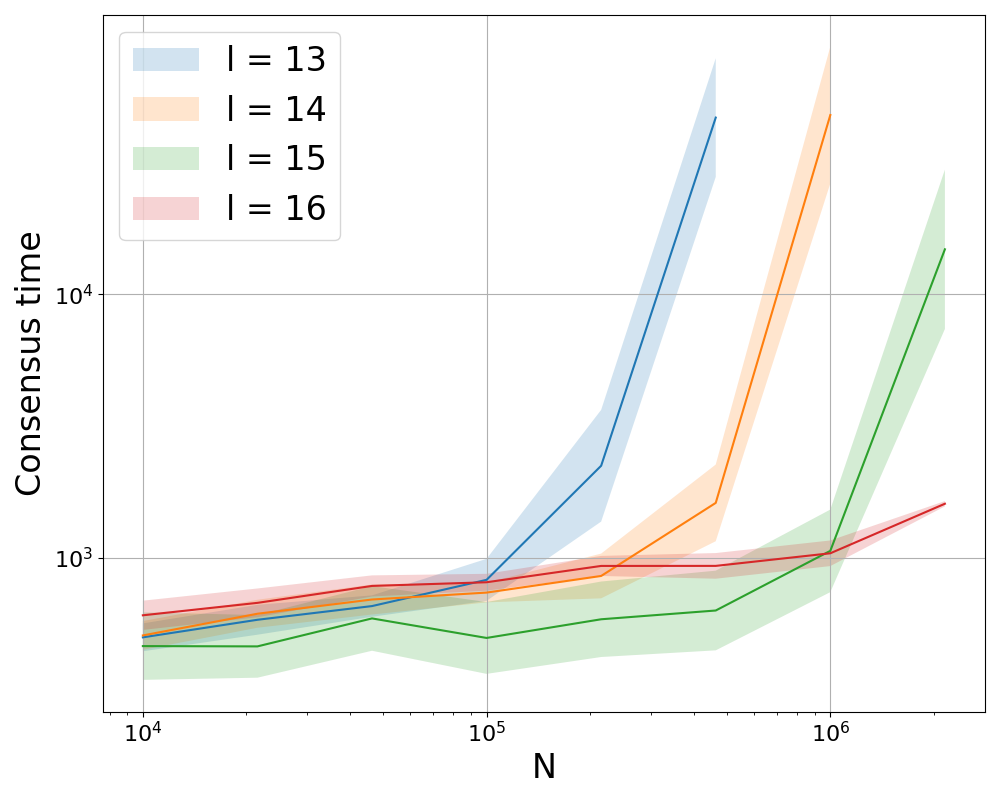}
                \caption{}
                \label{consensus_sim}
    \end{subfigure}
    \captionsetup{justification=raggedright,singlelinecheck=false}
    \caption{(a) PSD of the main frequency of $m$ (computed with the Fast Fourier Transform method) as a function of $l$, for both numerical solution and model simulations (fluctuations are negligible also in the latter case). 
    (b) Period of $m(t)$ (computed as the average distance between peaks) as a function of $l$, for both numerical solution and simulations. 
    (c) Minimum distance from the consensus state, $1-\max_{s\le t}|m(s)|$, achieved by the numerical solution as a function of $N$.
    (d) Time to reach the consensus state in simulations as a function of the population size $N$.}
\label{fig2}
\end{figure*}

%\subsection{Dynamical origin of the oscillations}
\medskip
\paragraph*{Dynamical origin of the oscillations.} To explain why the LVM shows an oscillating behavior, we focus on the flip probabilities of eq. \eqref{eq:5}. In the initial configuration with $m=0$ and no latent agents, the probabilities of having a flip to $+1$ or to $-1$ are equal. Random fluctuations then drive the evolution of the system. Suppose that at the first update, the selected agent flips from $-1$ to $+1$. This means $m$ increases by $2/N$ and $\lambda_+$ by $1/N$, while $\lambda_-$ is unchanged. As a result $\phi_{+-}$ decreases with respect to $\phi_{-+}$: we have $\Delta\phi=\phi_{-+}-\phi_{+-}\simeq 1/(2N)$. Hence, at the next update, the magnetization is more likely to increase than to go back to zero, $\lambda_+$ grows and $\Delta\phi$ increases as well. This mechanism determines a drift that does not exist in the original voter model, where the two flip probabilities are always equal. 

By iterating the above reasoning we see that the agents in latency tend to have the spin +1 (or more generally the same spin as the sign of the magnetization). However, after $l$ time steps, the agents that exit latency determine a decrease of $\lambda_+$ (higher than that of $\lambda_-$) and a gradual ri-equilibration of the flip probabilities. Since these agents are not frozen anymore and so can change opinion to $-1$, the growth of $m$ slows down. 
Once some of them flip, $\lambda_-$ starts growing and $\phi_{+-}$ becomes larger than $\phi_{-+}$ (\emph{i.e.}, $\Delta\phi$ becomes negative). 
We thus have the same situation just described but reverted towards the opposite sign of $m$. Opinion waves are hence due to agents with the same spin going into and exiting latency together.  

Such a dynamics is confirmed by Figure \ref{counterphase}, where we plot the numerical solutions for $\Delta\phi(t)$ and $m(t)$ at steady state (\emph{i.e.}, far from the initial fluctuations): these quantities oscillate with a quarter-period shift.

%\subsection{Parameters space}
\medskip
\paragraph*{Shape of the oscillations and asymptotic behavior.}

We finally investigate how the population size $N$ and the latency time $l$ influence the dynamics of the LVM.\footnote{The role of the initial magnetization $m_0$ is marginal as the initial configuration is quickly forgotten by the dynamics. Thus in simulations we set $m_0=0$ while to obtain the numerical solution we set $m_0=10^{-6}$.} 
%Specifically, we show how to turn on and off the opinion swings and in what cases our system is able to reach a consensus state ($m=\pm1$).
First, we characterize the frequency and amplitude of the oscillations of $m(t)$. 
Figure \ref{psd_cross} shows the Power Spectral Density (PSD) of the main frequency as a function of the latency time $l$.  Both simulations and the numerical solution show a sharp transition from a noisy behavior (where the magnetization remains close to zero) for values of $l \lesssim 3$, to a neat swinging phase with a specific main frequency -- whose PSD dominates over the others. 
Figure \ref{period_cross} shows instead the period of the oscillations. 
For $l\ge3$ this quantity grows almost linearly as a function of $l$. 
The agreement between simulations and the numerical solution is good until $l\simeq 10$ (see Supplementary Information S5).

%The time to reach a consensus requires a more accurate and detailed analysis, which we carried out by comparing the simulations with the numerical solution.  
Then we investigate whether the model dynamics is able to reach the consensus state. 
Figure \ref{iter_sol} shows that the numerical solution never reaches $|m|=1$: in the thermodynamic limit, opinions keep swinging forever. The maximum amplitude of the oscillations grows with $l$ and stabilizes around a value that is smaller than 1 but does not depend on $N$. 
Indeed, such local maxima of $m$ can be arbitrarily close to $1$ and be always compatible with eqs. \eqref{eq_m} and \eqref{eq_lam} (see Supplementary Information S6). 
Consensus is however reached in simulations as a finite-size effect. 
As Figure \ref{consensus_sim} shows, the time to consensus is rather short for small values of the population size $N$. However, for larger values of $N$ fluctuations become smaller and consensus time grows more than exponentially with population size. Furthermore, for fixed $N$, longer latencies $l$ are characterized by oscillations of higher amplitude that ease the reach of consensus.

%\section{Conclusions}
\medskip
\paragraph*{Conclusions.}
We have shown how the addition of a simple ingredient in the voter model, namely a constant latency time for agents after they change opinion, leads to the spontaneous emergence of deterministic dynamical oscillations in the average opinion. 
This behavior is totally different in nature both from the diffusive route to consensus of the original model and from the mean-reverting dynamics of noisy models that keep the system in a disordered state. 
To the best of our knowledge, we provided the first evidence of opinion swings in binary-choice models, which can help shed light on social contexts where the average opinion of the population features regular oscillatory patterns -- the emblematic example being the \emph{swing states} in the US political elections. 

A key to obtain oscillations in the LVM is the same latency time for each agent. Indeed the higher the variability of individual latencies the weaker the oscillations (see Supplementary Information S7); when latency times are completely random, we recover the setup by Lambiotte \emph{et al.} \cite{lambiotte2009}. 
The use of a common latency time for agents is however justified in contexts like political elections, particularly in the US where they take place every four years and the voting population is almost equally split into two opinions only.

The model can be generalized in many directions, for instance using different interaction structures of the population, including more than two opinion states, or different rules for the latency state. For instance, if we consider the alternative dynamic rule whereby agents who maintain their opinion enter latency (rather than those who change it), we get a model formulation in which the route to consensus is accelerated by a deterministic drift (see Supplementary Information S 8).

\newpage

\onecolumngrid

%\supplementary stuff
\setcounter{section}{0}
\renewcommand{\thesection}{S\arabic{section}}%
\setcounter{figure}{0}
\renewcommand{\thefigure}{S\arabic{figure}}%
\setcounter{equation}{0}
\renewcommand{\theequation}{S.\arabic{equation}}

\begin{center}
{\bf \large SUPPLEMENTARY INFORMATION}
\end{center}

\section{S1: Electoral results for all US states}

\begin{figure}[h]
    \centering
    \includegraphics[width=0.49\linewidth]{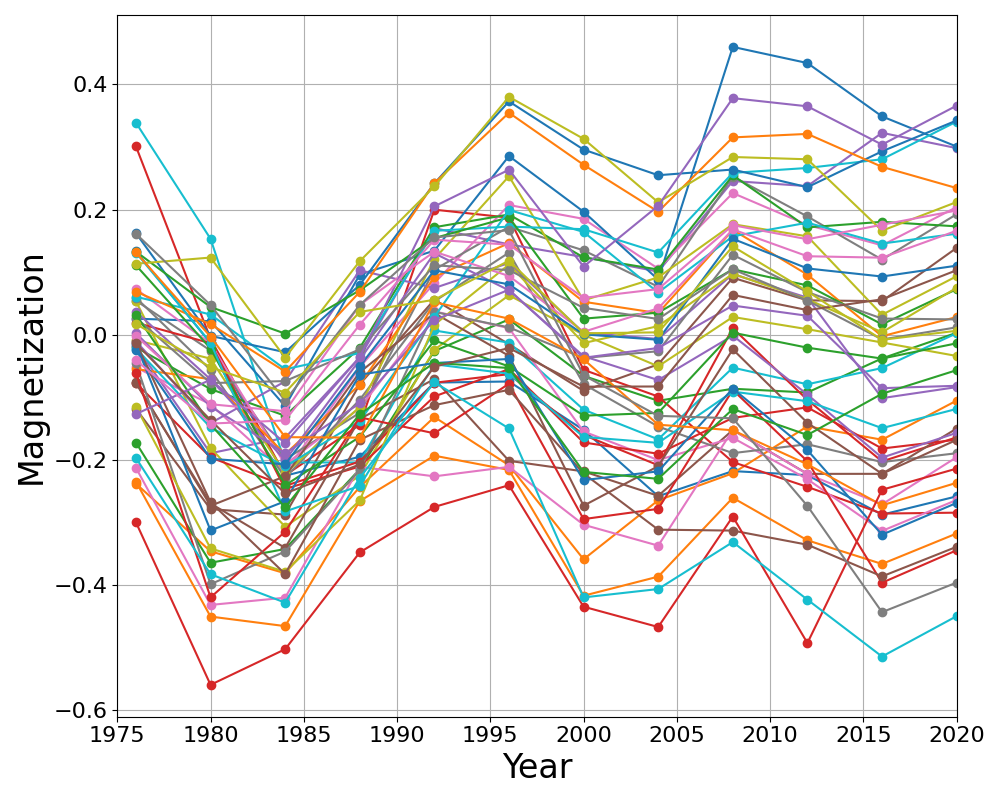}
    \caption{Oscillations in elections results for all US states (except DC). $+1$ represents a vote for Democrats and $-1$ for Republicans (votes for third parties are discarded). The so-called \emph{swing states} are those that oscillate around $m=0$ (and so they are crucial in determining the global outcome), yet all states in fact swing, following a common trend.}
    \label{swing}
\end{figure}

\section{S2: Model dynamics for all relevant variables}
\begin{figure*}[h]
    \begin{subfigure}[b]{0.49\textwidth}
        \centering
        \includegraphics[width=\linewidth]{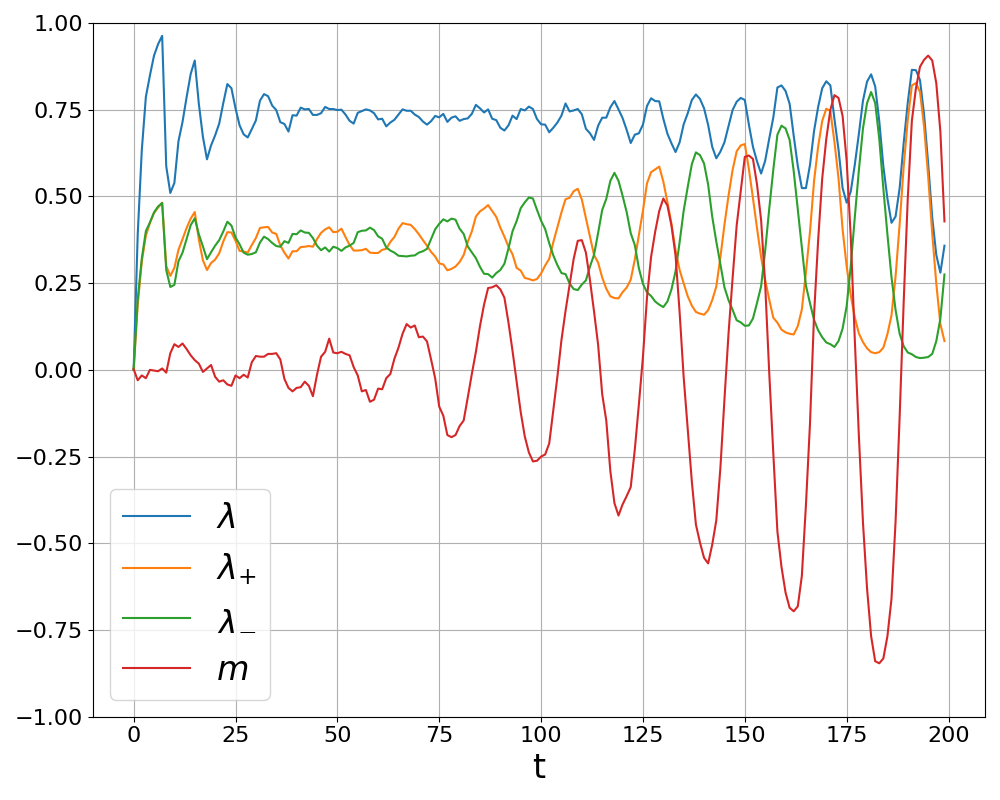}
                \caption{}
                \label{full1}
    \end{subfigure}
    \hfill
    \begin{subfigure}[b]{0.49\textwidth}
        \centering
        \includegraphics[width=\linewidth]{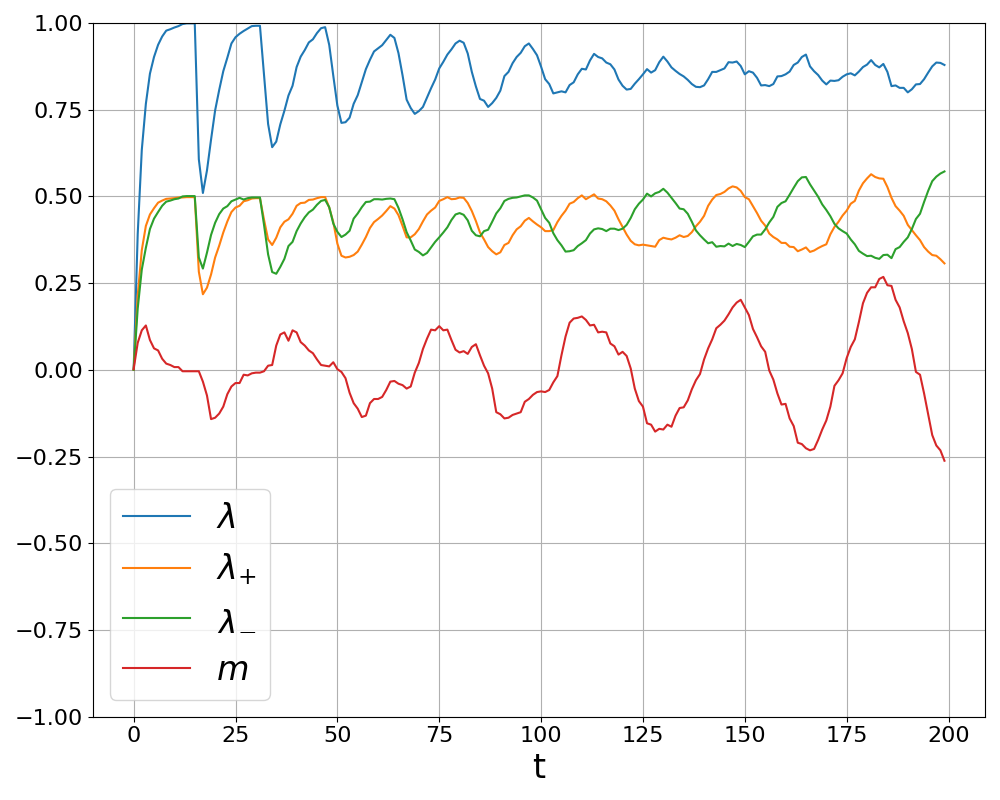}
                \caption{}
                \label{full2}
    \end{subfigure}
    \caption{Full plot of variables with $l=7$ (a) and $l=15$ (b).}
\label{fig1}
\end{figure*}

\section{S3: Approximate solution for small magnetization}

\begin{figure}[h]
    \centering
    \includegraphics[width=0.549\linewidth]{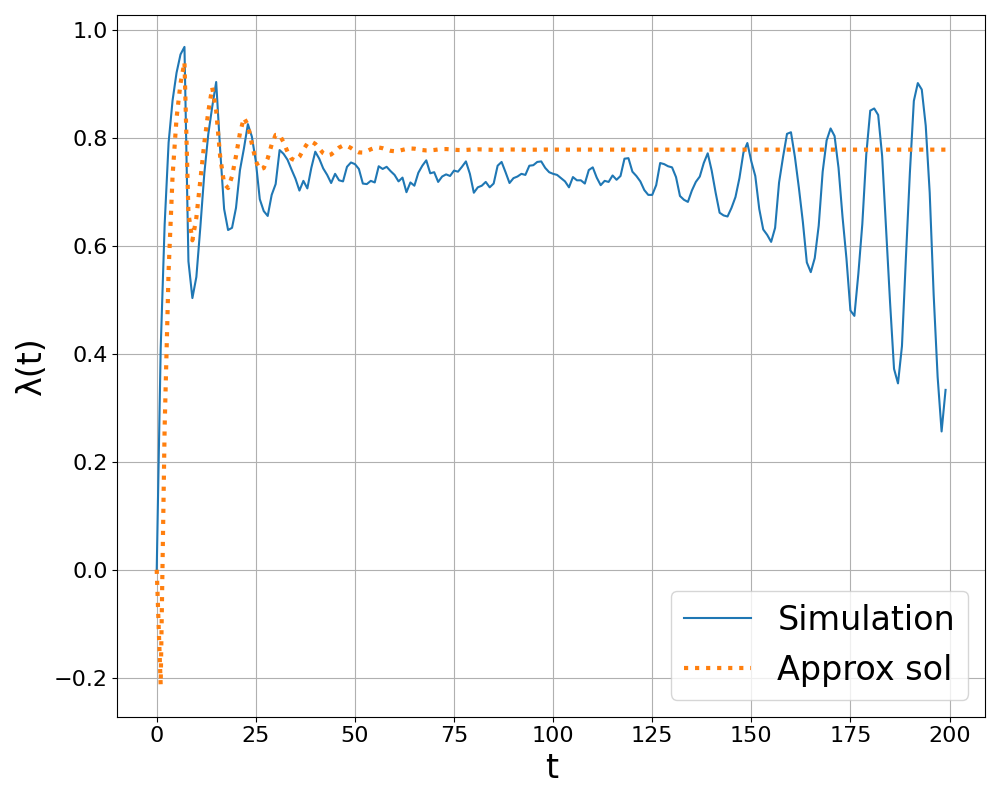}
    \caption{Simulation and approximate solution for $\lambda(t)$ have good overlap for a significant amount of time.}
    \label{approx}
\end{figure}

With respect to the system of DDEs describing the expected values of the main variables of the system (eqs 5 and 6 of the main text), it is interesting to analyse the behavior for small values of $m$, which leads to linear DDEs that can be solved with the step-method.
Let's suppose $m \approx 0$, then the solution for $E[\lambda(t)]$ for $t<nl$ with $n \in \mathbb{N} $ leads to
\begin{equation}\label{eq:10}
\lambda(t) = 1 - \sum_{k=0}^{n-1}\frac{1}{k!} \left( \frac{t-kl}{2} \right)^k e^{-\frac{t-kl}{2}}.
\end{equation}
This solution works well in predicting the actual evolution of the system as soon as the magnetization maintains a value around $0$, which is true for a significant amount of time, as can be inferred by fig. \ref{approx}. 
More importantly, it works as a proof of concept to understand the features and behavior of typical DDE solutions.
As common in these cases, the function obtained is not a smooth $\mathbb{C}^\infty$ function, but a piecewise one, as evident from \eqref{eq:10}. Indeed each term of the sum turns on until the $n_{th}$ one, where $t<nl$. The curve plotted in Figure \ref{approx} has $n=4$ and therefore correspond to the approximate solution (for $3l<t<4l$):
\begin{equation}
\lambda(t)=1 - e^ {-\frac{t}{2}} - \frac{1}{2} \left( \frac{t-2l}{2} \right)^2 e^{-\frac{t-2l}{2}} - \frac{1}{4} \left( \frac{t-3l}{3} \right)^3 e^{-\frac{t-3l}{3}}
\end{equation}
The typical solutions of DDEs are subject to smoothing \cite{bellman1963}. It can be in fact proven that \eqref{eq:10} has a discontinuity of its first derivative at $t=l$, a discontinuity of the second derivative at $t=2l$, of the third at $t=3l$ and so on, which makes the function smoother and smoother as time increases. Moreover, it can be easily proved that \eqref{eq:10} has a limit smaller than $1$ for $t \to +\infty$, since (being all the terms positive)
\begin{equation}
1 - e^ {-\frac{t}{2}} - \frac{1}{2} \left( \frac{t-2l}{2} \right)^2 e^{-\frac{t-2l}{2}} - ... < 1 -e^ {-\frac{t}{2}}
\end{equation}
and the right side of the equation asymptotically reaches $1$.
Yet this solution works fine in predicting the actual value of $\lambda$ only for a short time. When the oscillation of $m$ approaches a significant amplitude, $\lambda$ begins oscillating as well, as shown in Figure \ref{approx}.

%\newpage

\section{S4: The role of randomness on simulations}

\begin{figure}[h]
    \centering
    \includegraphics[width=0.49\linewidth]{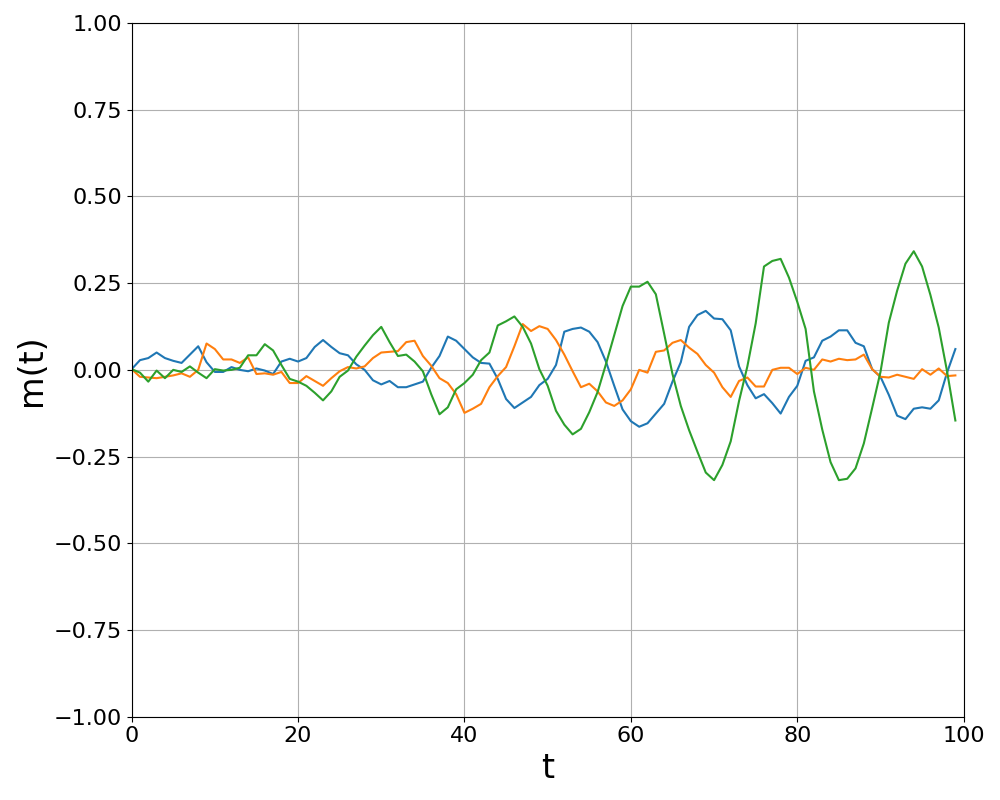}
    \caption{Simulations of the LVM dynamics with the same parameters ($N=1000$, $l=5$, $m_{0}=0$)  oscillate with the same period but have different phase and amplitude growth rates.}
    \label{randomness}
\end{figure}

While the behavior of the LVM is deterministic in nature and thus much more predictable than the original voter model, randomness still plays an important role in determining the evolution of individual realizations of the system. 
Indeed, while the period of the oscillations depends on the value of the latency time, fluctuations set the growth rate of the amplitude and the phase of $m(t)$ (see Figure \ref{randomness}). 
Therefore it is not possible to compare the evolution of simulations with the expected value of $m(t)$ from eq (5) of the main text.

\section{S5: Measurements of period and Power Spectral Density}

\begin{figure}[h]
    \centering
    \includegraphics[width=0.49\linewidth]{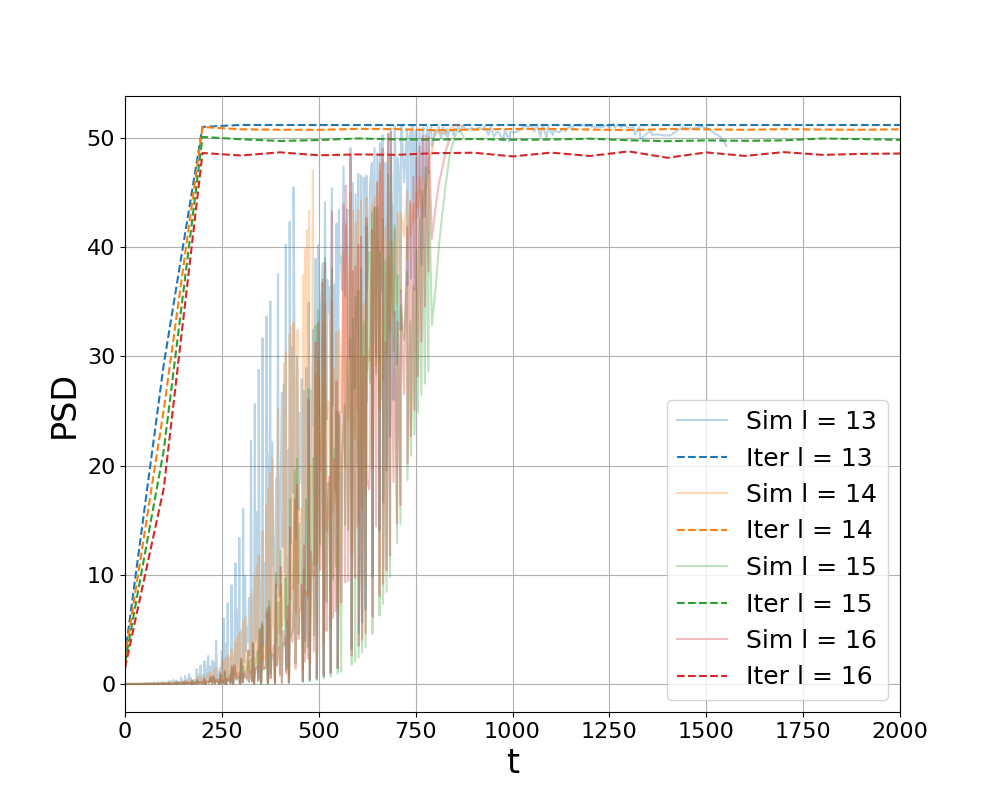}\hfill
    \includegraphics[width=0.49\linewidth]{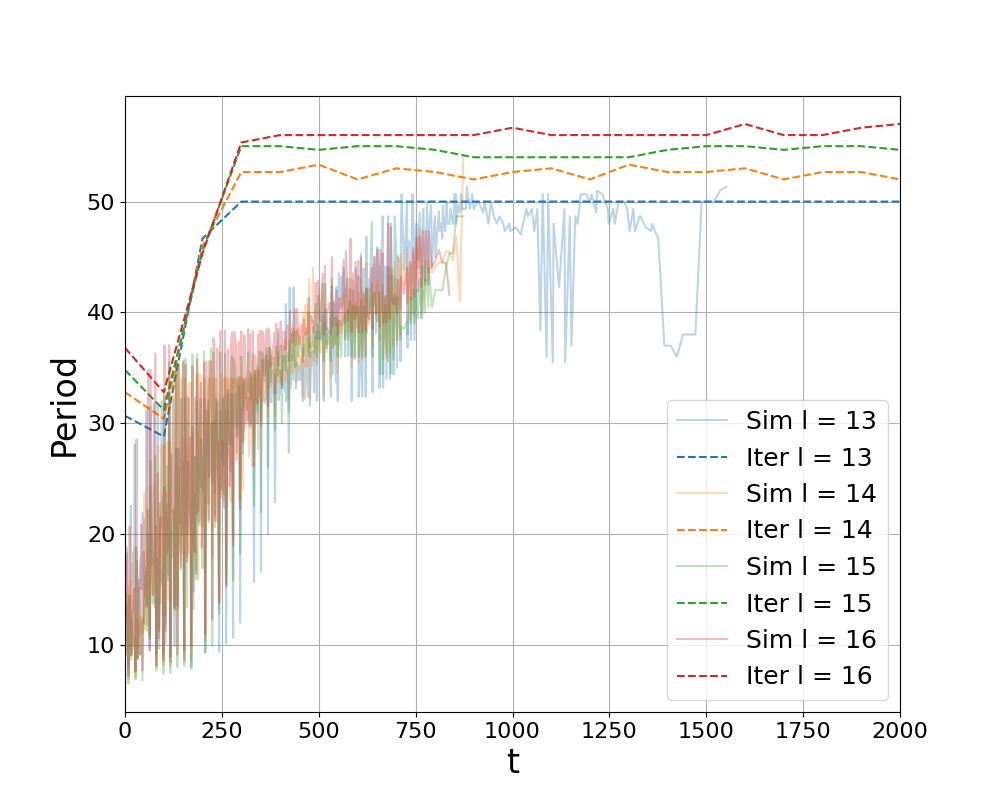}
    \caption{PSD and period for simulations and numerical solutions ($N=10000$), computed on a moving time window of width $100$.}
    \label{psd_period}
\end{figure}

To detect the period of the oscillations and the PSD of the main frequency of $m(t)$, we used the average distance between peaks and the Fast Fourier Transform (FFT) of $m(t)$, respectively.
While results of the PSD always agree between simulation and numerical solution, this is not the case for the period in the region $l\gtrsim10$. This happens because both quantities stabilise to their stationary values after a transient region, which lasts much longer in simulations (see Figure \ref{psd_period}). 
We sample after the transient to obtain reliable measurements (those reported in Figure 2 of the main text), however for $l\gtrsim10$ the period is not able to reach a stable value before consensus is reached because of random fluctuations.

%\newpage

\section{S6: Existence of local maxima of the magnetization smaller than $1$}

In the main text, we provide evidence that the model dynamics can reach the consensus state only as a finite-size effect. 
Here we show that the model admits local maxima of $m$ that are arbitrarily close to consensus. 
To this end we rewrite our equations in terms of $\Delta\lambda(t)=\lambda_{+}(t)-\lambda_{-}(t)$. We get:
\begin{equation}
    m'(t)=\lambda(t)\left[\frac{\Delta\lambda(t)}{\lambda(t)}-m(t)\right]
\end{equation}
The stationary points of $m$ are thus defined by $m=\Delta\lambda/\lambda$,\footnote{The fact that the solution of DDEs is not a $\mathbb{C}^\infty$ function is not a limitation, because it only has a corner at $t=l$ and then gets smoother as time goes by, so we can rely on the condition on the first derivative to find the stationary points.} 
a condition that can be rewritten as 
\begin{equation}
m(t)=1-\frac{2\lambda_{-}(t)}{\lambda(t)}\label{max}
\end{equation}
As $m$ grows towards $+1$, also $\lambda_{-}$ decreases, so that $m$ eventually reaches a value satisfying eq.\eqref{max} and smaller than 1. 
This is not an inflection point because $m'$ changes sign there. 
We can conclude that the solution of the model equations are compatible with the existence of local maxima at an arbitrarily small distance from $1$. 
The same argument of course applies to the opposite case $m=-1$.

\section{S7: Heterogeneous latency times}

\begin{figure}[h]
    \centering
    \includegraphics[width=0.49\linewidth]{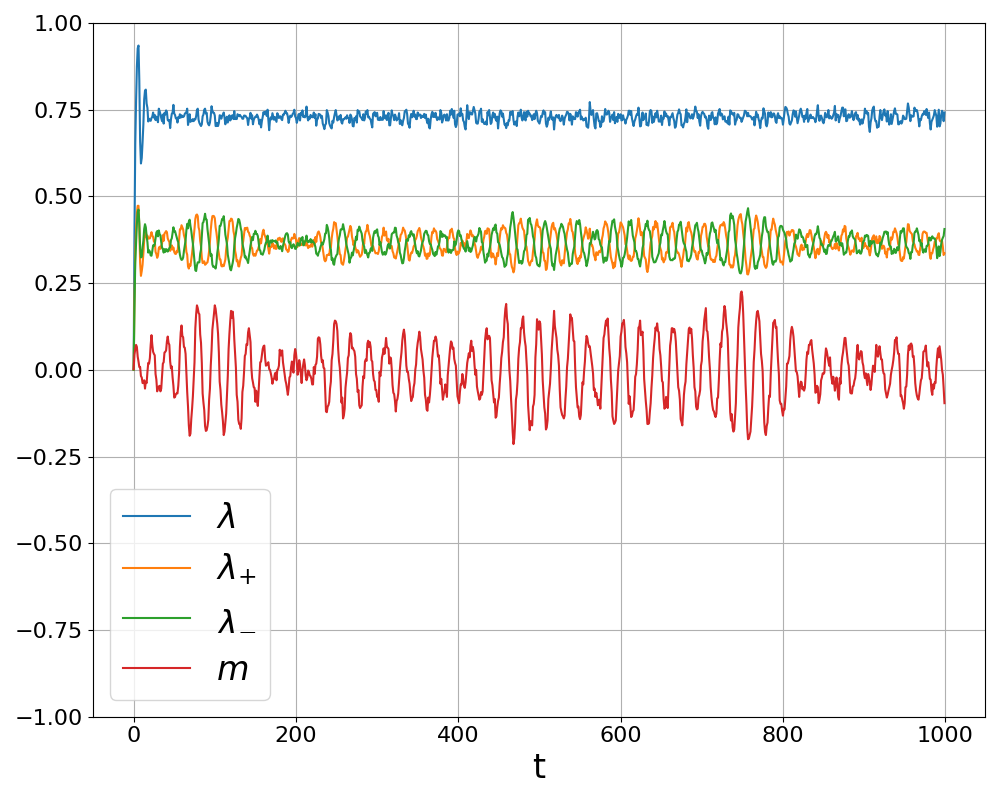}
    \caption{Simulation with same parameters used in Figure \ref{full1} but with agents having heterogeneous latency times.}
    \label{differential}
\end{figure}

Here we investigate what happens when the latency time is not the same for all agents. 
Figure \ref{differential} shows what happens in the same configuration of Figure \ref{full1} but agents have heterogeneous latencies, which are distributed as a Gaussian centered at $l=7$ and $\sigma=1$. 
The dynamics become fuzzy, oscillations have higher frequencies and the growth in amplitude changes drastically. 
The broader the distribution of latencies the noisier the dynamics. For very heterogeneous latencies we retrieve the behavior observed in \cite{lambiotte2009}, where the magnetisation remains around zero. 
This confirms that the swinging behaviour emerges due to the synchronisation of nodes going into and exiting latency together.
On the contrary, opinion swings persist when we put some agents in the latent state at the beginning of the simulation. 

\section{S8: Opposite model}

\begin{figure}[h]
    \centering
    \includegraphics[width=0.49\linewidth]{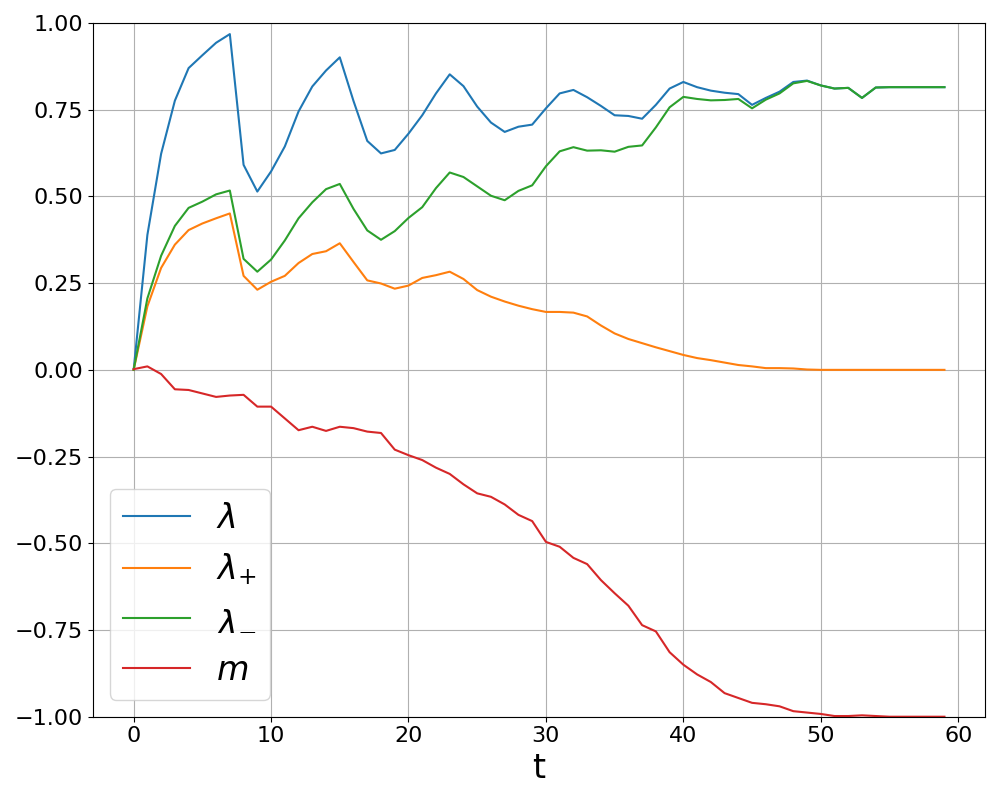}
    \caption{Single run of the opposite LVM dynamics, with $N=1000$ and $l=7$ (the same setup used in Figure \ref{full1}).}
    \label{opposite}
\end{figure}

Here we study the ``opposite'' model where agents go in latency when they are chosen for the update but not flip.
According to the discussion provided in the main text on the origin of oscillations in the LVM, such an opposite model should not oscillate. 
Indeed the evolution in the first $l$ time-steps is the same in the two versions of the model. 
At this point, $m$ has grown (say towards $+1$) and it is more likely to pick for the update an agent with $s=+1$. 
Since there are also more neighbors with $s=+1$, this agent is more likely to remain in her state than to flip, and so it goes into latency. 
Overall, agents with the same spin of the sign of $m$ are likely to be in the latent state; they cannot change opinion and therefore the upward trend of $m$ is locked (it does not go back to $0$). 
Thus in the opposite model, latency gives rise to a drift that quickly pushes the system to the consensus state. This is confirmed by numerical simulations, reported in Figure \ref{opposite}.

\begin{thebibliography}{29}%
\makeatletter
\providecommand \@ifxundefined [1]{%
 \@ifx{#1\undefined}
}%
\providecommand \@ifnum [1]{%
 \ifnum #1\expandafter \@firstoftwo
 \else \expandafter \@secondoftwo
 \fi
}%
\providecommand \@ifx [1]{%
 \ifx #1\expandafter \@firstoftwo
 \else \expandafter \@secondoftwo
 \fi
}%
\providecommand \natexlab [1]{#1}%
\providecommand \enquote  [1]{``#1''}%
\providecommand \bibnamefont  [1]{#1}%
\providecommand \bibfnamefont [1]{#1}%
\providecommand \citenamefont [1]{#1}%
\providecommand \href@noop [0]{\@secondoftwo}%
\providecommand \href [0]{\begingroup \@sanitize@url \@href}%
\providecommand \@href[1]{\@@startlink{#1}\@@href}%
\providecommand \@@href[1]{\endgroup#1\@@endlink}%
\providecommand \@sanitize@url [0]{\catcode `\\12\catcode `\$12\catcode
  `\&12\catcode `\#12\catcode `\^12\catcode `\_12\catcode `\%12\relax}%
\providecommand \@@startlink[1]{}%
\providecommand \@@endlink[0]{}%
\providecommand \url  [0]{\begingroup\@sanitize@url \@url }%
\providecommand \@url [1]{\endgroup\@href {#1}{\urlprefix }}%
\providecommand \urlprefix  [0]{URL }%
\providecommand \Eprint [0]{\href }%
\providecommand \doibase [0]{https://doi.org/}%
\providecommand \selectlanguage [0]{\@gobble}%
\providecommand \bibinfo  [0]{\@secondoftwo}%
\providecommand \bibfield  [0]{\@secondoftwo}%
\providecommand \translation [1]{[#1]}%
\providecommand \BibitemOpen [0]{}%
\providecommand \bibitemStop [0]{}%
\providecommand \bibitemNoStop [0]{.\EOS\space}%
\providecommand \EOS [0]{\spacefactor3000\relax}%
\providecommand \BibitemShut  [1]{\csname bibitem#1\endcsname}%
\let\auto@bib@innerbib\@empty
%</preamble>
\bibitem [{\citenamefont {Castellano}\ \emph {et~al.}(2009)\citenamefont
  {Castellano}, \citenamefont {Fortunato},\ and\ \citenamefont
  {Loreto}}]{castellano2009}%
  \BibitemOpen
  \bibfield  {author} {\bibinfo {author} {\bibfnamefont {C.}~\bibnamefont
  {Castellano}}, \bibinfo {author} {\bibfnamefont {S.}~\bibnamefont
  {Fortunato}},\ and\ \bibinfo {author} {\bibfnamefont {V.}~\bibnamefont
  {Loreto}},\ }\bibfield  {title} {\bibinfo {title} {Statistical physics of
  social dynamics},\ }\href {https://doi.org/10.1103/RevModPhys.81.591}
  {\bibfield  {journal} {\bibinfo  {journal} {Reviews of Modern Physics}\
  }\textbf {\bibinfo {volume} {81}},\ \bibinfo {pages} {591} (\bibinfo {year}
  {2009})}\BibitemShut {NoStop}%
\bibitem [{\citenamefont {Galam}(2012)}]{galam2012}%
  \BibitemOpen
  \bibfield  {author} {\bibinfo {author} {\bibfnamefont {S.}~\bibnamefont
  {Galam}},\ }\href {https://doi.org/10.1007/978-1-4614-2032-3} {\emph
  {\bibinfo {title} {Sociophysics : a physicist's modeling of psycho-political
  phenomena}}},\ Springer complexity\ (\bibinfo  {publisher} {Springer New
  York},\ \bibinfo {address} {New York},\ \bibinfo {year} {2012})\BibitemShut
  {NoStop}%
\bibitem [{\citenamefont {Redner}(2019)}]{redner2019}%
  \BibitemOpen
  \bibfield  {author} {\bibinfo {author} {\bibfnamefont {S.}~\bibnamefont
  {Redner}},\ }\bibfield  {title} {\bibinfo {title} {Reality-inspired voter
  models: A mini-review},\ }\href
  {https://doi.org/https://doi.org/10.1016/j.crhy.2019.05.004} {\bibfield
  {journal} {\bibinfo  {journal} {Comptes Rendus Physique}\ }\textbf {\bibinfo
  {volume} {20}},\ \bibinfo {pages} {275} (\bibinfo {year} {2019})}\BibitemShut
  {NoStop}%
\bibitem [{\citenamefont {Peralta}\ \emph {et~al.}(2022)\citenamefont
  {Peralta}, \citenamefont {Kert{\'e}sz},\ and\ \citenamefont
  {I{\~n}iguez}}]{peralta2022}%
  \BibitemOpen
  \bibfield  {author} {\bibinfo {author} {\bibfnamefont {A.~F.}\ \bibnamefont
  {Peralta}}, \bibinfo {author} {\bibfnamefont {J.}~\bibnamefont
  {Kert{\'e}sz}},\ and\ \bibinfo {author} {\bibfnamefont {G.}~\bibnamefont
  {I{\~n}iguez}},\ }\href@noop {} {\bibinfo {title} {Opinion dynamics in social
  networks: From models to data}} (\bibinfo {year} {2022}),\ \Eprint
  {https://arxiv.org/abs/2201.01322} {arXiv:2201.01322} \BibitemShut {NoStop}%
\bibitem [{\citenamefont {Clifford}\ and\ \citenamefont
  {Sudbury}(1973)}]{clifford1973}%
  \BibitemOpen
  \bibfield  {author} {\bibinfo {author} {\bibfnamefont {P.}~\bibnamefont
  {Clifford}}\ and\ \bibinfo {author} {\bibfnamefont {A.}~\bibnamefont
  {Sudbury}},\ }\bibfield  {title} {\bibinfo {title} {{A model for spatial
  conflict}},\ }\href {https://doi.org/10.1093/biomet/60.3.581} {\bibfield
  {journal} {\bibinfo  {journal} {Biometrika}\ }\textbf {\bibinfo {volume}
  {60}},\ \bibinfo {pages} {581} (\bibinfo {year} {1973})}\BibitemShut
  {NoStop}%
\bibitem [{\citenamefont {Holley}\ and\ \citenamefont
  {Liggett}(1975)}]{holley1975}%
  \BibitemOpen
  \bibfield  {author} {\bibinfo {author} {\bibfnamefont {R.~A.}\ \bibnamefont
  {Holley}}\ and\ \bibinfo {author} {\bibfnamefont {T.~M.}\ \bibnamefont
  {Liggett}},\ }\bibfield  {title} {\bibinfo {title} {{Ergodic Theorems for
  Weakly Interacting Infinite Systems and the Voter Model}},\ }\href
  {https://doi.org/10.1214/aop/1176996306} {\bibfield  {journal} {\bibinfo
  {journal} {The Annals of Probability}\ }\textbf {\bibinfo {volume} {3}},\
  \bibinfo {pages} {643 } (\bibinfo {year} {1975})}\BibitemShut {NoStop}%
\bibitem [{\citenamefont {Schelling}(1973)}]{schelling1973}%
  \BibitemOpen
  \bibfield  {author} {\bibinfo {author} {\bibfnamefont {T.~C.}\ \bibnamefont
  {Schelling}},\ }\bibfield  {title} {\bibinfo {title} {Hockey helmets,
  concealed weapons, and daylight saving: A study of binary choices with
  externalities},\ }\href {http://www.jstor.org/stable/173406} {\bibfield
  {journal} {\bibinfo  {journal} {The Journal of Conflict Resolution}\ }\textbf
  {\bibinfo {volume} {17}},\ \bibinfo {pages} {381} (\bibinfo {year}
  {1973})}\BibitemShut {NoStop}%
\bibitem [{\citenamefont {Asch}(2003)}]{asch2003}%
  \BibitemOpen
  \bibfield  {author} {\bibinfo {author} {\bibfnamefont {S.~E.}\ \bibnamefont
  {Asch}},\ }\bibfield  {title} {\bibinfo {title} {Effects of group pressure
  upon the modification and distortion of judgments},\ }in\ \href@noop {}
  {\emph {\bibinfo {booktitle} {Organizational influence processes}}}\
  (\bibinfo  {publisher} {Routledge},\ \bibinfo {year} {2003})\ pp.\ \bibinfo
  {pages} {295--303}\BibitemShut {NoStop}%
\bibitem [{\citenamefont {Redner}(2001)}]{redner2001}%
  \BibitemOpen
  \bibfield  {author} {\bibinfo {author} {\bibfnamefont {S.}~\bibnamefont
  {Redner}},\ }\href {https://doi.org/10.1017/CBO9780511606014} {\emph
  {\bibinfo {title} {A Guide to First-Passage Processes}}}\ (\bibinfo
  {publisher} {Cambridge University Press},\ \bibinfo {year}
  {2001})\BibitemShut {NoStop}%
\bibitem [{\citenamefont {Holme}\ and\ \citenamefont
  {Liljeros}(2015)}]{holme2015}%
  \BibitemOpen
  \bibfield  {author} {\bibinfo {author} {\bibfnamefont {P.}~\bibnamefont
  {Holme}}\ and\ \bibinfo {author} {\bibfnamefont {F.}~\bibnamefont
  {Liljeros}},\ }\bibfield  {title} {\bibinfo {title} {Mechanistic models in
  computational social science},\ }\href
  {https://doi.org/https://doi.org/10.3389/fphy.2015.00078} {\bibfield
  {journal} {\bibinfo  {journal} {Frontiers in Physics}\ }\textbf {\bibinfo
  {volume} {3}},\ \bibinfo {pages} {78} (\bibinfo {year} {2015})}\BibitemShut
  {NoStop}%
\bibitem [{\citenamefont {Galesic}\ and\ \citenamefont
  {Stein}(2019)}]{galesic2019}%
  \BibitemOpen
  \bibfield  {author} {\bibinfo {author} {\bibfnamefont {M.}~\bibnamefont
  {Galesic}}\ and\ \bibinfo {author} {\bibfnamefont {D.~L.}\ \bibnamefont
  {Stein}},\ }\bibfield  {title} {\bibinfo {title} {Statistical physics models
  of belief dynamics: Theory and empirical tests},\ }\href
  {https://doi.org/https://doi.org/10.1016/j.physa.2018.12.011} {\bibfield
  {journal} {\bibinfo  {journal} {Physica A: Statistical Mechanics and its
  Applications}\ }\textbf {\bibinfo {volume} {519}},\ \bibinfo {pages} {275}
  (\bibinfo {year} {2019})}\BibitemShut {NoStop}%
\bibitem [{\citenamefont {Moscovici}(1980)}]{moscovici1980}%
  \BibitemOpen
  \bibfield  {author} {\bibinfo {author} {\bibfnamefont {S.}~\bibnamefont
  {Moscovici}},\ }\bibfield  {title} {\bibinfo {title} {Toward a theory of
  conversion behavior},\ }in\ \href
  {https://doi.org/https://doi.org/10.1016/S0065-2601(08)60133-1} {\emph
  {\bibinfo {booktitle} {Advances in experimental social psychology}}},\
  Vol.~\bibinfo {volume} {13}\ (\bibinfo  {publisher} {Elsevier},\ \bibinfo
  {year} {1980})\ pp.\ \bibinfo {pages} {209--239}\BibitemShut {NoStop}%
\bibitem [{\citenamefont {Moscovici}(1985)}]{moscovici1985}%
  \BibitemOpen
  \bibfield  {author} {\bibinfo {author} {\bibfnamefont {S.}~\bibnamefont
  {Moscovici}},\ }\bibinfo {title} {Innovation and minority influence},\ in\
  \href {https://doi.org/10.1017/CBO9780511897566.003} {\emph {\bibinfo
  {booktitle} {Perspectives on Minority Influence}}},\ \bibinfo {series and
  number} {European Studies in Social Psychology},\ \bibinfo {editor} {edited
  by\ \bibinfo {editor} {\bibfnamefont {E.~v.}\ \bibnamefont {Avermaet}},
  \bibinfo {editor} {\bibfnamefont {G.}~\bibnamefont {Mugny}},\ and\ \bibinfo
  {editor} {\bibfnamefont {S.}~\bibnamefont {Moscovici}}}\ (\bibinfo
  {publisher} {Cambridge University Press},\ \bibinfo {address} {Cambridge},\
  \bibinfo {year} {1985})\ pp.\ \bibinfo {pages} {9--52}\BibitemShut {NoStop}%
\bibitem [{\citenamefont {Mobilia}\ \emph {et~al.}(2007)\citenamefont
  {Mobilia}, \citenamefont {Petersen},\ and\ \citenamefont
  {Redner}}]{mobilia2007}%
  \BibitemOpen
  \bibfield  {author} {\bibinfo {author} {\bibfnamefont {M.}~\bibnamefont
  {Mobilia}}, \bibinfo {author} {\bibfnamefont {A.}~\bibnamefont {Petersen}},\
  and\ \bibinfo {author} {\bibfnamefont {S.}~\bibnamefont {Redner}},\
  }\bibfield  {title} {\bibinfo {title} {On the role of zealotry in the voter
  model},\ }\href {https://doi.org/10.1088/1742-5468/2007/08/P08029} {\bibfield
   {journal} {\bibinfo  {journal} {Journal of Statistical Mechanics: Theory and
  Experiment}\ }\textbf {\bibinfo {volume} {2007}},\ \bibinfo {pages} {P08029}
  (\bibinfo {year} {2007})}\BibitemShut {NoStop}%
\bibitem [{\citenamefont {Galam}\ and\ \citenamefont
  {Jacobs}(2007)}]{galam2007}%
  \BibitemOpen
  \bibfield  {author} {\bibinfo {author} {\bibfnamefont {S.}~\bibnamefont
  {Galam}}\ and\ \bibinfo {author} {\bibfnamefont {F.}~\bibnamefont {Jacobs}},\
  }\bibfield  {title} {\bibinfo {title} {The role of inflexible minorities in
  the breaking of democratic opinion dynamics},\ }\href
  {https://doi.org/https://doi.org/10.1016/j.physa.2007.03.034.} {\bibfield
  {journal} {\bibinfo  {journal} {Physica A: Statistical Mechanics and its
  Applications}\ }\textbf {\bibinfo {volume} {381}},\ \bibinfo {pages} {366}
  (\bibinfo {year} {2007})}\BibitemShut {NoStop}%
\bibitem [{\citenamefont {Xie}\ \emph {et~al.}(2011)\citenamefont {Xie},
  \citenamefont {Sreenivasan}, \citenamefont {Korniss}, \citenamefont {Zhang},
  \citenamefont {Lim},\ and\ \citenamefont {Szymanski}}]{xie2011}%
  \BibitemOpen
  \bibfield  {author} {\bibinfo {author} {\bibfnamefont {J.}~\bibnamefont
  {Xie}}, \bibinfo {author} {\bibfnamefont {S.}~\bibnamefont {Sreenivasan}},
  \bibinfo {author} {\bibfnamefont {G.}~\bibnamefont {Korniss}}, \bibinfo
  {author} {\bibfnamefont {W.}~\bibnamefont {Zhang}}, \bibinfo {author}
  {\bibfnamefont {C.}~\bibnamefont {Lim}},\ and\ \bibinfo {author}
  {\bibfnamefont {B.~K.}\ \bibnamefont {Szymanski}},\ }\bibfield  {title}
  {\bibinfo {title} {Social consensus through the influence of committed
  minorities},\ }\href {https://doi.org/10.1103/PhysRevE.84.011130} {\bibfield
  {journal} {\bibinfo  {journal} {Physical Review E}\ }\textbf {\bibinfo
  {volume} {84}},\ \bibinfo {pages} {011130} (\bibinfo {year}
  {2011})}\BibitemShut {NoStop}%
\bibitem [{\citenamefont {Stark}\ \emph {et~al.}(2008)\citenamefont {Stark},
  \citenamefont {Tessone},\ and\ \citenamefont {Schweitzer}}]{stark2008}%
  \BibitemOpen
  \bibfield  {author} {\bibinfo {author} {\bibfnamefont {H.-U.}\ \bibnamefont
  {Stark}}, \bibinfo {author} {\bibfnamefont {C.~J.}\ \bibnamefont {Tessone}},\
  and\ \bibinfo {author} {\bibfnamefont {F.}~\bibnamefont {Schweitzer}},\
  }\bibfield  {title} {\bibinfo {title} {Decelerating microdynamics can
  accelerate macrodynamics in the voter model},\ }\href
  {https://doi.org/10.1103/PhysRevLett.101.018701} {\bibfield  {journal}
  {\bibinfo  {journal} {Physical Review Letters}\ }\textbf {\bibinfo {volume}
  {101}},\ \bibinfo {pages} {018701} (\bibinfo {year} {2008})}\BibitemShut
  {NoStop}%
\bibitem [{\citenamefont {Wang}\ \emph {et~al.}(2014)\citenamefont {Wang},
  \citenamefont {Liu}, \citenamefont {Wang},\ and\ \citenamefont
  {Zhang}}]{wang2014}%
  \BibitemOpen
  \bibfield  {author} {\bibinfo {author} {\bibfnamefont {Z.}~\bibnamefont
  {Wang}}, \bibinfo {author} {\bibfnamefont {Y.}~\bibnamefont {Liu}}, \bibinfo
  {author} {\bibfnamefont {L.}~\bibnamefont {Wang}},\ and\ \bibinfo {author}
  {\bibfnamefont {Y.}~\bibnamefont {Zhang}},\ }\bibfield  {title} {\bibinfo
  {title} {Freezing period strongly impacts the emergence of a global consensus
  in the voter model},\ }\href {https://doi.org/10.1038/srep03597} {\bibfield
  {journal} {\bibinfo  {journal} {Scientific Reports}\ }\textbf {\bibinfo
  {volume} {4}},\ \bibinfo {pages} {3597} (\bibinfo {year} {2014})}\BibitemShut
  {NoStop}%
\bibitem [{\citenamefont {Lambiotte}\ \emph {et~al.}(2009)\citenamefont
  {Lambiotte}, \citenamefont {Saram\"aki},\ and\ \citenamefont
  {Blondel}}]{lambiotte2009}%
  \BibitemOpen
  \bibfield  {author} {\bibinfo {author} {\bibfnamefont {R.}~\bibnamefont
  {Lambiotte}}, \bibinfo {author} {\bibfnamefont {J.}~\bibnamefont
  {Saram\"aki}},\ and\ \bibinfo {author} {\bibfnamefont {V.~D.}\ \bibnamefont
  {Blondel}},\ }\bibfield  {title} {\bibinfo {title} {Dynamics of latent
  voters},\ }\href {https://doi.org/10.1103/PhysRevE.79.046107} {\bibfield
  {journal} {\bibinfo  {journal} {Physical Review E}\ }\textbf {\bibinfo
  {volume} {79}},\ \bibinfo {pages} {046107} (\bibinfo {year}
  {2009})}\BibitemShut {NoStop}%
\bibitem [{\citenamefont {Huo}\ and\ \citenamefont
  {Durrett}(2018)}]{huo2018latent}%
  \BibitemOpen
  \bibfield  {author} {\bibinfo {author} {\bibfnamefont {R.}~\bibnamefont
  {Huo}}\ and\ \bibinfo {author} {\bibfnamefont {R.}~\bibnamefont {Durrett}},\
  }\bibfield  {title} {\bibinfo {title} {Latent voter model on locally
  tree-like random graphs},\ }\href
  {https://doi.org/https://doi.org/10.1016/j.spa.2017.08.004} {\bibfield
  {journal} {\bibinfo  {journal} {Stochastic Processes and their Applications}\
  }\textbf {\bibinfo {volume} {128}},\ \bibinfo {pages} {1590} (\bibinfo {year}
  {2018})}\BibitemShut {NoStop}%
\bibitem [{\citenamefont {Montes~de Oca}\ \emph {et~al.}(2011)\citenamefont
  {Montes~de Oca}, \citenamefont {Ferrante}, \citenamefont {Scheidler},
  \citenamefont {Pinciroli}, \citenamefont {Birattari},\ and\ \citenamefont
  {Dorigo}}]{montesdeoca2011}%
  \BibitemOpen
  \bibfield  {author} {\bibinfo {author} {\bibfnamefont {M.~A.}\ \bibnamefont
  {Montes~de Oca}}, \bibinfo {author} {\bibfnamefont {E.}~\bibnamefont
  {Ferrante}}, \bibinfo {author} {\bibfnamefont {A.}~\bibnamefont {Scheidler}},
  \bibinfo {author} {\bibfnamefont {C.}~\bibnamefont {Pinciroli}}, \bibinfo
  {author} {\bibfnamefont {M.}~\bibnamefont {Birattari}},\ and\ \bibinfo
  {author} {\bibfnamefont {M.}~\bibnamefont {Dorigo}},\ }\bibfield  {title}
  {\bibinfo {title} {Majority-rule opinion dynamics with differential latency:
  a mechanism for self-organized collective decision-making},\ }\href
  {https://doi.org/10.1007/s11721-011-0062-z} {\bibfield  {journal} {\bibinfo
  {journal} {Swarm Intelligence}\ }\textbf {\bibinfo {volume} {5}},\ \bibinfo
  {pages} {305} (\bibinfo {year} {2011})}\BibitemShut {NoStop}%
\bibitem [{\citenamefont {Scheidler}(2011)}]{scheidler2011}%
  \BibitemOpen
  \bibfield  {author} {\bibinfo {author} {\bibfnamefont {A.}~\bibnamefont
  {Scheidler}},\ }\bibfield  {title} {\bibinfo {title} {Dynamics of majority
  rule with differential latencies},\ }\href
  {https://doi.org/10.1103/PhysRevE.83.031116} {\bibfield  {journal} {\bibinfo
  {journal} {Physical Review E}\ }\textbf {\bibinfo {volume} {83}},\ \bibinfo
  {pages} {031116} (\bibinfo {year} {2011})}\BibitemShut {NoStop}%
\bibitem [{\citenamefont {Fern{\'a}ndez-Gracia}\ \emph
  {et~al.}(2014)\citenamefont {Fern{\'a}ndez-Gracia}, \citenamefont {Suchecki},
  \citenamefont {Ramasco}, \citenamefont {San~Miguel},\ and\ \citenamefont
  {Egu{\'\i}luz}}]{fernandezgracia2014}%
  \BibitemOpen
  \bibfield  {author} {\bibinfo {author} {\bibfnamefont {J.}~\bibnamefont
  {Fern{\'a}ndez-Gracia}}, \bibinfo {author} {\bibfnamefont {K.}~\bibnamefont
  {Suchecki}}, \bibinfo {author} {\bibfnamefont {J.}~\bibnamefont {Ramasco}},
  \bibinfo {author} {\bibfnamefont {M.}~\bibnamefont {San~Miguel}},\ and\
  \bibinfo {author} {\bibfnamefont {V.~M.}\ \bibnamefont {Egu{\'\i}luz}},\
  }\bibfield  {title} {\bibinfo {title} {Is the voter model a model for
  voters?},\ }\href {https://doi.org/10.1103/PhysRevLett.112.158701} {\bibfield
   {journal} {\bibinfo  {journal} {Physical Review Letters}\ }\textbf {\bibinfo
  {volume} {112}},\ \bibinfo {pages} {158701} (\bibinfo {year}
  {2014})}\BibitemShut {NoStop}%
\bibitem [{\citenamefont {Braha}\ and\ \citenamefont
  {de~Aguiar}(2017)}]{braha2017}%
  \BibitemOpen
  \bibfield  {author} {\bibinfo {author} {\bibfnamefont {D.}~\bibnamefont
  {Braha}}\ and\ \bibinfo {author} {\bibfnamefont {M.~A.~M.}\ \bibnamefont
  {de~Aguiar}},\ }\bibfield  {title} {\bibinfo {title} {Voting contagion:
  Modeling and analysis of a century of u.s. presidential elections},\ }\href
  {https://doi.org/10.1371/journal.pone.0177970} {\bibfield  {journal}
  {\bibinfo  {journal} {PLOS ONE}\ }\textbf {\bibinfo {volume} {12}},\ \bibinfo
  {pages} {e0177970} (\bibinfo {year} {2017})}\BibitemShut {NoStop}%
\bibitem [{\citenamefont {Kononovicius}(2017)}]{kononovicius2017}%
  \BibitemOpen
  \bibfield  {author} {\bibinfo {author} {\bibfnamefont {A.}~\bibnamefont
  {Kononovicius}},\ }\bibfield  {title} {\bibinfo {title} {Empirical analysis
  and agent-based modeling of the lithuanian parliamentary elections},\ }\href
  {https://doi.org/10.1155/2017/7354642} {\bibfield  {journal} {\bibinfo
  {journal} {Complexity}\ }\textbf {\bibinfo {volume} {2017}},\ \bibinfo
  {pages} {7354642} (\bibinfo {year} {2017})}\BibitemShut {NoStop}%
\bibitem [{\citenamefont {Ansolabehere}\ \emph {et~al.}(2006)\citenamefont
  {Ansolabehere}, \citenamefont {Rodden},\ and\ \citenamefont
  {Snyder}}]{ansolabhere2006}%
  \BibitemOpen
  \bibfield  {author} {\bibinfo {author} {\bibfnamefont {S.}~\bibnamefont
  {Ansolabehere}}, \bibinfo {author} {\bibfnamefont {J.}~\bibnamefont
  {Rodden}},\ and\ \bibinfo {author} {\bibfnamefont {J.~M.}\ \bibnamefont
  {Snyder}},\ }\bibfield  {title} {\bibinfo {title} {Purple america},\ }\href
  {https://doi.org/10.1257/jep.20.2.97} {\bibfield  {journal} {\bibinfo
  {journal} {The Journal of Economic Perspectives}\ }\textbf {\bibinfo {volume}
  {20}},\ \bibinfo {pages} {97} (\bibinfo {year} {2006})}\BibitemShut {NoStop}%
\bibitem [{\citenamefont {Atkinson}\ \emph {et~al.}(2021)\citenamefont
  {Atkinson}, \citenamefont {Coggins}, \citenamefont {Stimson},\ and\
  \citenamefont {Baumgartner}}]{atkinson2021dynamics}%
  \BibitemOpen
  \bibfield  {author} {\bibinfo {author} {\bibfnamefont {M.~L.}\ \bibnamefont
  {Atkinson}}, \bibinfo {author} {\bibfnamefont {K.~E.}\ \bibnamefont
  {Coggins}}, \bibinfo {author} {\bibfnamefont {J.~A.}\ \bibnamefont
  {Stimson}},\ and\ \bibinfo {author} {\bibfnamefont {F.~R.}\ \bibnamefont
  {Baumgartner}},\ }\href@noop {} {\emph {\bibinfo {title} {The Dynamics of
  Public Opinion}}}\ (\bibinfo  {publisher} {Cambridge University Press},\
  \bibinfo {year} {2021})\BibitemShut {NoStop}%
\bibitem [{\citenamefont {Data}\ and\ \citenamefont
  {Lab}(2017)}]{DVN/42MVDX_2017}%
  \BibitemOpen
  \bibfield  {author} {\bibinfo {author} {\bibfnamefont {M.~E.}\ \bibnamefont
  {Data}}\ and\ \bibinfo {author} {\bibfnamefont {S.}~\bibnamefont {Lab}},\
  }\bibfield  {title} {\bibinfo {title} {{U.S. President 1976–2020}},\
  }\bibfield  {journal} {\bibinfo  {journal} {Harvard Dataverse}\ }\href
  {https://doi.org/10.7910/DVN/42MVDX} {10.7910/DVN/42MVDX} (\bibinfo {year}
  {2017})\BibitemShut {NoStop}%
\bibitem [{\citenamefont {Bellman}\ and\ \citenamefont
  {Cooke}(1963)}]{bellman1963}%
  \BibitemOpen
  \bibfield  {author} {\bibinfo {author} {\bibfnamefont {R.~E.}\ \bibnamefont
  {Bellman}}\ and\ \bibinfo {author} {\bibfnamefont {K.~L.}\ \bibnamefont
  {Cooke}},\ }\href@noop {} {\emph {\bibinfo {title} {Differential-Difference
  Equations}}}\ (\bibinfo  {publisher} {Academic Press},\ \bibinfo {address}
  {Santa Monica, CA},\ \bibinfo {year} {1963})\BibitemShut {NoStop}%
\end{thebibliography}
\end{document}